# CMS ECAL VFE design, production and testing

W. Lustermann[a,1], D. Abadjiev[b], G. Dissertori[a], M. Dejardin[c], T. Gadek[a], L.T. Martin[b], and K. Stachon[a] on behalf of the CMS collaboration

[a] *ETH Zurich,*
  *ETH Hönggerberg, Institute for Particle Physics and Astrophysics, Otto-Stern-Weg 5, CH-8093 Zurich, Switzerland*
[b] *Northeastern University,*
  *Department of Physics, 360 Huntington Ave, Boston, Ma 02115, USA*
[c] *IRFU, CEA, Université Paris-Saclay,*
  *F-91191 Gif Sur Yvette, France*
  E-mail: `Werner.Lustermann@cern.ch`

ABSTRACT: Maintaining the required performance of the CMS electromagnetic calorimeter (ECAL) barrel at the High-Luminosity Large Hadron Collider (HL-LHC) requires the replacement of the entire on-detector electronics. 12240 new very front end (VFE) cards will amplify and digitize the signals of 62100 lead-tungstate crystals instrumented with avalanche photodiodes. The VFE cards host five channels of CATIA pre-amplifier ASICs followed by LiTE-DTU ASICs, which digitize signals with 160MS/s and 12bit resolution. We present the strategy and infrastructure developed for achieving the required reliability of less than 0.5% failing channels over the expected lifetime of 20 years. This includes the choice of standards, design for reliability and manufacturing, as well as factory acceptance tests, reception testing, environmental stress screening and calibration of the VFE cards.

# 1. Introduction

The electromagnetic calorimeter (ECAL) barrel [1] of the Compact Muon Solenoid (CMS) experiment [2] is made of 61200 lead-tungstate crystals read out by Avalanche Photodiodes (APDs). The ECAL barrel readout electronics will be upgraded for HL-LHC. The upgraded readout electronics are arranged into towers of 5x5 channels, comprising of five Very Front End (VFE) cards, one digital interface card (FE), and one Low Voltage Regulator (LVR) card conditioning the power of one tower. Each VFE hosts five identical channels of a trans-impedance amplifier (CATIA) followed by an analog-to-digital converter and data transmission unit (LiTE-DTU). The tower electronics will be replaced for operation at HL-LHC [3] for various purposes:

- Coping with the increased level 1 trigger latency of 12.5µs, presently ~3µs.
- Mitigation the effects of a ~six-fold increase in radiation dose, such as the increasing leakage current in the APDs.
- Enabling time reconstruction with a precision of <30ps for energies larger than 50GeV.
- Introduction of data streaming to the back-end electronics, enabling flexibility in the level 1 trigger primitives' generation and rejection of so-called spike events, stemming from direct nuclear interactions in the junction region of the APDs.

This involves the production and testing of ~80000 CATIA [4] and LiTE-DTU [5] ASICs and the assembly and testing of ~14600 VFE cards.

Key features of the new ECAL readout are the higher bandwidth (35MHz) of the CATIA transimpedance amplifier, two gain outputs (gains: 1 and 10), a four times higher sampling rate of 160MS/s of the 12bit samples, lossless data compression by the LiTE-DTU, and streaming of all data with 1.28GS/s to the FE card and further to the back-end electronics.

# 2. Materials and Methods

## 2.1 VFE reliability requirements

By the end of LHC phase 1, occurring at the end of 2025, the presently installed VFE cards will have operated for ~20 years since their production. The new VFE electronics will be produced in 2024. They are required to work until the end of 2041 according to the HL-LHC schedule [6]. Adding three years of contingency yields an expected lifetime of 20 years.

The ECAL barrel is a single layer detector. Missing channels directly impact energy and timing resolution, and missing readout towers permit electrons and photons to escape unidentified. Repair or maintenance of the electronics during phase 2 of HL-LHC operation is practically impossible, because it would require a shutdown of ~26 months to replace electronics in all 36 supermodules. Hence, it is not foreseen. Since ECAL is well performing with ~1% of broken channels and since the VFEs are not the only source of failures, we require the number of failing VFE channels to be less than 0.5% at the end-of-life after 20 years.

### 2.1.1 Design for reliability and manufacturing

The VFE card was designed with reliability (design for reliability - DFR) and manufacturing (design for manufacturing - DFM) in mind. A first important step is the selection of the appropriate standard for the Printed Circuit Board (PCB) layout (see for example [7] for the selection of the IPC standard). We require IPC class 3, following the IPC-6011/6012 standard for rigid PCBs. Class 3 is appropriate in our application because maintenance or replacement of individual failing VFE cards is not possible for the entire lifetime. The increased cost for fabrication and assembly of class 3 PCBs vs class 2 PCBs, typically 10% to 20%, is relatively low and acceptable. This said, we consider IPC class 3/A, which includes requirements for military avionics and the space industry, as exaggerated and the related higher cost increase as unjustified.

Moreover, we require:

- Surface finishing of Electric Nickel-Gold (ENIG), ensuring good solderability of the PCBs.
- Core and prepreg materials with glass transition temperature (TG) >170ºC.
- Impedance control of transmission lines to ±10%, tuned to standard materials and checked by means of test coupons on a sample's bases.
- High speed signal guarding using VIA stitching, see Figure 1 for an example.
- Exclusive use of through hole VIAs.
- Standard copper layer thickness of 35µm.
- PCB fabrication in panels, arranged in collaboration with the PCB assembler.

Furthermore, we have chosen relatively large components, 0603 wherever possible and 0402 otherwise, easing rework. We select automotive grade components (AEC-200) or better and chose SMD connectors over through hole connectors with the exception of the SAMTEC connector which is mandatory for compatibility with the legacy system. Connector pins are gold plated. Components must be provided on reels or in Electrostatic Discharge (ESD) safe trays. We apply appropriate derating of components, <80% for voltage and current and <50% for power. Higher voltage parts shall be coated, avoiding the growth of dendrites. See Figure 2 for the coating of the decoupling capacitor rated for 630V (nominal operation voltage: <500V).

**2.1.2 Production for reliability**

As for the PCB itself, we require IPC class 3 following the IPC-A-610 standard for the PCB assembly. In class 3, we find more stringent criteria on solder joints. The placement of components must be more precise, and through hole filling is at minimum 75% compared to 50% for class 2. Figure 3 shows an example of an incompletely filled through hole.

We specify a standard lead-free solder paste and require the bake-out of components prior to assembly. In order to avoid systematic failures on large parts of the production, factory acceptance tests (FATs) are mandatory, in particular automatic optical inspection (AOI) and electrical testing of all completed VFE cards. The test results allow a continuous monitoring of the first pass yield (FPY). This allows for spotting shifts in the production quality early on. We require the manufacturer to stop production and consult with us should the FPY fall below the threshold of 92%. Similarly, the overall production yield (including reworked parts) shall be monitored. If it drops below 95% the manufacturer shall consult with us.

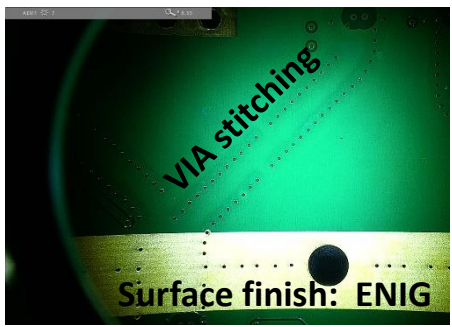

Figure 1: Via stitching and surface finish implemented on the VFE card.

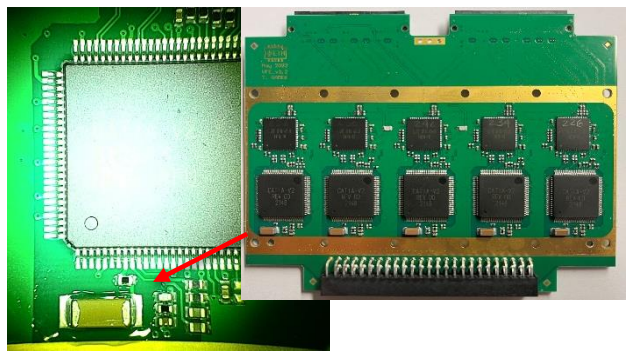

Figure 2: Coating of 630V rated decoupling capacitor.

Moreover, the manufacturer shall be well qualified for the required assembly work: production and components handling shall happen in an ESD safe environment and components shall be managed using a dedicated components management system. The manufacturer shall be ISO 9001 certified (quality assurance and control) and the personal shall be trained for PCB assembly following IPC-A-610, as well as for rework and repair following IPC 7711/7721.

Finally, the manufacturer shall use processes and materials compliant with laws and regulations for the protection of the environment. In particular, RoHS compliant components shall be used and the manufacturer shall be ISO 14001 certified.

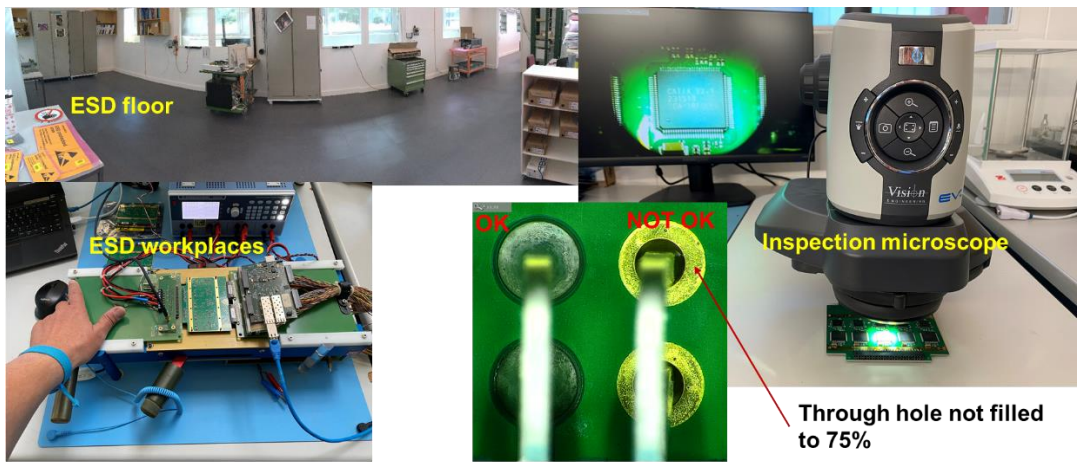

Figure 3: Images of the antistatic floor (top left) in our electronics laboratory, ESD workplace (bottom left), PCB inspection microscope (right) and insufficiently filled through hole (bottom middle).

## 2.2 VFE production and testing

Another important aspect of achieving high quality and reliability is systematic reception testing and environmental stress screening of all VFE cards in an ESD safe environment. For this, we have added an ESD floor to our laboratory and fitted our workplaces with antistatic mats and bracelets. Moreover, we invested in a PCB inspection microscope (see Figure 3).

### 2.2.1 Production plan

The VFE production starts with a pre-production batch of ~400 cards. The large quantity allows an in-depth qualification of the chosen assembly company. We reserve three months for completing the qualification of the pre-production. All cards will be electrically tested in a setup identical to the one used during the FAT. A subset of ~40 cards will undergo environmental stress screening providing us with the failure rate of the VFE cards as a function of time. This allows for an estimation of the expected failure rate after 20 years of operation. All remaining cards will be installed in the ECAL spare super-module for a large-scale system test.

The qualification of the pre-production is followed by production in batches of 500 cards per week, corresponding to a total of ~29 weeks. This enables us to follow the production pace with our reception tests. The assembly company will deliver the batch as soon as production and FATs are complete, once per week. We will perform electrical testing of the batch during one week, followed by one week of environmental stress screening and one week of calibration. In this way, there are always 4 batches in the production and qualification pipeline: one is produced, one is electrically tested, one is environmental stress screened, and one is calibrated.

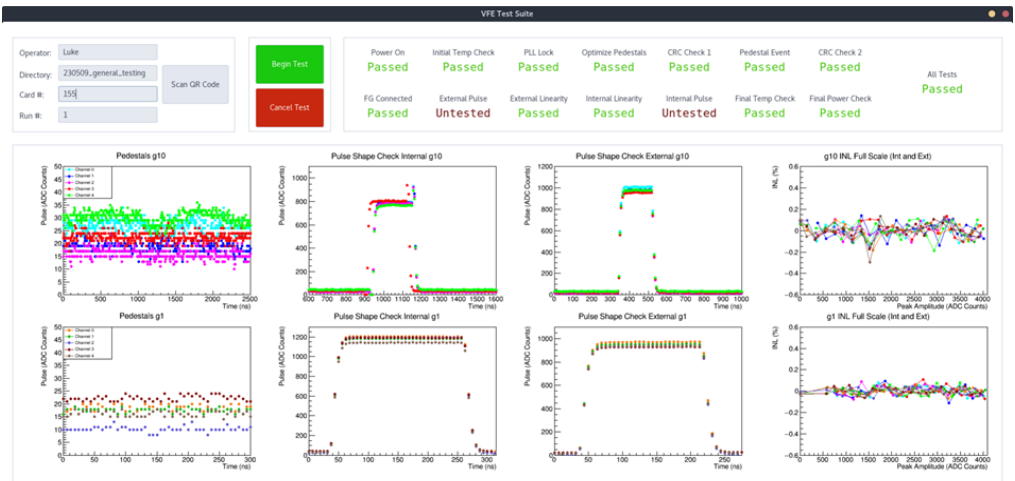

Figure 4: VFE test-setup graphical user interface.

### 2.2.2 Electrical test and calibration setup

We developed a test and calibration setup for the VFE cards. The setup comprises of a laboratory power supply, an adapter PCB for power connections and test pulse injection, a VFE Interface Converting Electronics (VICE++) card [8], and a laptop. The VICE++ enables the configuration of the CATIA and LiTE-DTU registers from the laptop. Moreover, it hosts data buffers allowing the storage of a snapshot of data sampled at 160 MS/s for later transmission to the laptop by means of electrical gigabit ethernet links. A typical test sequence comprises of:

- Scanning all LiTE-DTU phased lock loop (PLL) configuration settings, which provides the PLL locking voltage range, and selecting the PLL settings.
- Calibrating the LiTE-DTU ADCs via a reference voltage provided by the CATIA.
- Scanning all CATIA pedestal DAC values and measurement of the corresponding pedestals. Selection of the pedestal DAC values for subsequent measurements.
- Measurement of the pedestal values.
- Measurement of test-pulses generated by the CATIAs' internal test-pulse generators.

The software features a graphical user interface (Figure 4), providing main test and pass/fail information and displaying plots of the pedestal and test-pulse measurements for both CATIA gains. At the end of each test, the results are stored in a database hosted by CERN's database on demand service. Calibration of the VFE card gains is achieved with the same setup and an additional external precision test pulse generator. We use a Tektronix AFG31252, 250MHz, 14bit, 2GS/s arbitrary function generator. Figure 5 shows an example linearity measurement. The setup allows the cross calibration of all ECAL readout channels with a single source prior to their installation into CMS as well as the cross calibration of the test pulse generators integrated into each CATIA.

### 2.2.3 Environmental stress screening

Environmental stress screening (ESS) allows assessment of the reliability of the VFE card by measuring the failure rate versus time. We expect to see three phases: first, an infantile mortality phase with decreasing failure rate; second, a useful life phase with almost constant failure rate; and last, a wear out phase with increasing failure rate. We will measure the failure rate curve with ~40 preproduction cards by active thermal cycling. The number of power cycles of the VFE cards will be approximately ten per year when in use, resulting in a total of 200 power cycles for 20 years of operation. The VFE cards will be tested together with the LVR cards. This ensures the proper powering of the VFE cards during the test as well as a realistic load for the LVR cards. The cards will be installed in test boxes (see Figure 6). Heating is achieved by the power dissipation of the VFE and LVR cards themselves. Temperature is cycled between ambient air temperature and 70°C approximately once per hour. Numerous ventilators, some exchanging the air in the box for ambient air, and others stirring the air inside the box, controlled by an Arduino microcontroller, provide a uniform temperature of 70±1.5°C. Keithley DAQ 6510 instruments equipped with Keithley 7702 switch cards are used to monitor the temperatures on the VFE and LVR cards. A total of 12 boxes will host up to 108 LVR and 540 VFE cards, providing the necessary capacity to follow the production of 500 VFE cards per week. Safety of the setup will be provided by a Siemens PLC system reading two additional PT1000 temperature sensors per box. During production, we will apply the same procedure to VFE and LVR cards, with a number of cycles to age them just beyond the end of infantile mortality.

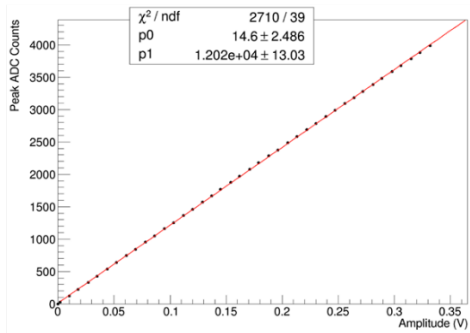

Figure 5: Example linearity measurement with external test pulse generator.

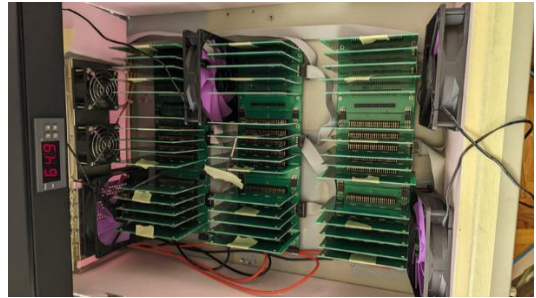

Figure 6: ESS box for 45 VFE and 9 LVR cards with ventilators for ambient air exchange (left) as well as ventilators for internal air steering. Stable temperatures of 70 ±1.5°C were achieved.

## 3. Summary and conclusion

Maintenance and reliability are crucial elements for the lifetime of components. In the case of the ECAL VFE card, maintenance is excluded, and consequently we concentrate our effort on obtaining the desired reliability. This is achieved by a series of measures: the choice of appropriate standards, DFR, DFM, components selection, qualification of the PCB assembler, FATs, and by extensive in-house reception and qualification testing, including electrical testing and ESS.

**Acknowledgments**

The authors thank Alexandre Dolgopolov and Alexander Singovski from University of Notre Dame, USA, for providing the VICE++ cards used in the VFE test setup. Furthermore, we thank Predrag Milenovic, Milorad Mijic and Nikola Rasevic from Belgrade University and Lazar Cokic from CERN for the development of the ESS setup. This research was supported by Swiss National Science Foundation (SNF) via the grant SNF FLARE 201476.